\documentclass[twocolumn]{aastex62}

\newcommand{\arcs}{$^{\prime\prime}$} % Arcseconds
\newcommand{\beq}{\begin{equation}\begin{aligned}}
\newcommand{\eeq}{\end{aligned}\end{equation}}

\newcommand{\msun}{M$_\odot$}

\usepackage{scalerel,amssymb}
\usepackage{amsmath}
\usepackage{enumitem}
\usepackage[]{url}
\usepackage{lineno}
\usepackage{CJKutf8}
\usepackage[caption=false]{subfig}
\defcitealias{karachentsev2013}{K13}
\defcitealias{kourkchi2017}{KT17}

\accepted{\today}
\submitjournal{ApJ}

\shorttitle{Isolated Dwarfs in the Local Volume}
\shortauthors{Carlsten et al.}

\begin{document}
\begin{CJK*}{UTF8}{gbsn}

\title{ELVES-Field: Isolated Dwarf Galaxy Quenched Fractions Rise Below $M_* \approx 10^7$~\msun }

\correspondingauthor{Scott G. Carlsten}
\email{scarlsten@gmail.com}

\author[0000-0002-5382-2898]{Scott G. Carlsten}
\affil{Department of Astrophysical Sciences, 4 Ivy Lane, Princeton University, Princeton, NJ 08540}

\author[0000-0001-9592-4190]{Jiaxuan Li (李嘉轩)}
\affiliation{Department of Astrophysical Sciences, 4 Ivy Lane, Princeton University, Princeton, NJ 08540}

\author[0000-0002-5612-3427]{Jenny E. Greene}
\affil{Department of Astrophysical Sciences, 4 Ivy Lane, Princeton University, Princeton, NJ 08540}

\author[0000-0001-8251-933X]{Alex Drlica-Wagner}
\affiliation{Fermi National Accelerator Laboratory, P.O.\ Box 500, Batavia, IL 60510, USA}
\affiliation{Kavli Institute for Cosmological Physics, University of Chicago, Chicago, IL 60637, USA}
\affiliation{Department of Astronomy and Astrophysics, University of Chicago, Chicago, IL 60637, USA}

\author[0000-0002-1841-2252]{Shany Danieli}
% \altaffiliation{Hubble Fellow}
% \affil{Department of Astrophysical Sciences, 4 Ivy Lane, Princeton University, Princeton, NJ 08540}
% \affiliation{Department of Astrophysical Sciences, 4 Ivy Lane, Princeton University, Princeton, NJ 08540, USA}
\affiliation{School of Physics and Astronomy, Tel Aviv University, Tel Aviv 69978, Israel}

\begin{abstract}
We use a new sample of low-mass ($M_* < 10^9$~\msun) isolated galaxies from the Exploration of Local VolumE Survey - Field (ELVES-Field) to examine the star formation properties and sizes of field dwarf galaxies in the Local Volume (LV; $D<10$ Mpc). This volume-limited sample was selected from nearly 3,000 square degrees of imaging, relying on surface brightness fluctuations to determine distances to the majority of the systems and is complete to $M_* \approx 10^6$~\msun. Across the surveyed area, we catalog over 2300 candidate LV dwarfs, of which we confirm 95 as genuine LV members and reject over 1600 as background contaminants, with the remaining 600 candidates still requiring a distance measurement. Of the confirmed LV dwarfs, 46 are either new discoveries or confirmed via a distance measurement for the first time here. We explore different environmental criteria to select isolated dwarfs but primarily focus on dwarfs that are $>2\times R_{\mathrm{vir}}$ in projection from any known group with $M_\star > 10^9$ \msun. We find that, at higher dwarf masses ($M_\star \gtrsim 10^7$ \msun), essentially all field dwarfs are star-forming as has been found before. In contrast, at $M_\star \lesssim 10^7$ \msun, $\sim30\%$ of field dwarfs appear to be quenched. Finally, we find that isolated dwarfs are noticeably smaller ($\sim 20\%$) than satellite dwarfs of the same stellar mass, regardless of quenched status.

\end{abstract}

\section{Introduction}

Dwarf galaxies are the most abundant galaxies in the Universe and also among the most fragile. Stellar feedback \citep{larson1974}, environmental processing \citep{grebel2003, mayer2006, putman2021}, and even reionization \citep{bullock2000} can all substantially alter the star-formation history of dwarf galaxies, depending on their mass and larger-scale environment. Ideally, one would want a large sample of completely isolated dwarf galaxies spanning a wide mass range to study their star-formation histories without the confounding influence of larger halos. Indeed, it is now well-established that satellite dwarfs around MW-like hosts show a steeply rising quenched fraction below $\lesssim 10^8$ \msun \; in stellar mass, reaching near unity below $\lesssim 10^6$ \msun \;\citep{wetzel2015, weisz2015, simon2019, putman2021, carlsten2022, greene2023, geha2024}. This observational result generally agrees with the predictions from hydrodynamic simulations \citep{simpson2018, akins2021, font2021b, samuel2022, engler2023, pan2023, vannest2023, rodriguez_cardoso2025}. 

Spectroscopic surveys like the Sloan Digital Sky Survey \citep[SDSS;][]{sdss7} and the Galaxy and Mass Assembly Survey \citep[GAMA;][]{driver2011, driver2022} are sufficiently complete down to masses $M_* \gtrsim 10^8$~\msun, and previous work has investigated the quenched fraction down to this completeness limit. In particular, using SDSS, \citet{geha2012} show that galaxies with stellar mass $ M_* \lesssim 10^{9.5}$~\msun\ are always star-forming unless they are neighbors with a massive $M_* > 10^{10}$~\msun\ companion. 

It may be that at lower stellar mass, however, isolated dwarfs exhibit a rise in quenched fraction. There are several known field dwarfs above the ultra-faint dwarf \citep[UFD, $M_\star\lesssim10^5$ \msun;][]{simon2019} mass scale that appear to be quenched \citep[e.g.][]{makarov2012, karachentsev2015_kks3, polzin2021, li2024_hedgehog, sand2024}. Additionally, multiple recent theoretical results (both simulation-based and semi-analytic) predict appreciable ($\sim 30\%$) quenched fractions in the $M_\star \sim 10^6-10^7$ \msun\; range \citep{christensen2024, kim2024, wang2024}. While the physical mechanism thought to quench low-mass satellites \citep[e.g. ram-pressure stripping;][]{fillingham2015, zhu2024} would not be active in the field, reionization is expected to quench even isolated dwarfs below some mass threshold \citep{bullock2000, benson2002, rey2020}. Reionization quenching is generally thought to be important primarily at the mass-scale of UFDs and less so in the $M_\star \sim 10^6-10^8$\msun\; range \citep{okamoto2009}, although it may still play a role. Additionally, stellar feedback might lead to self-quenching of isolated dwarfs \citep{samuel2022} or dwarf-dwarf mergers might lead to gas removal and quenching without the need for a massive host \citep{kadofong2024}. 

The observed quenched fraction for field dwarfs is thus a key unknown in the study of dwarf galaxies and one that requires a complete, volume-limited sample of field dwarfs to infer. Fortunately, large samples of dwarf galaxies have proliferated in the past decade with the advent of large imaging surveys. While historically most dwarf research focused on the Local Group \citep[e.g. see][and references therein]{mateo1998, koposov2008, mcconnachie2012, simon2019, drlica2020}, recent years have seen a large growth in samples down to ``classical'' dwarf masses of $M_* \approx 10^6$~\msun\ \citep[e.g][]{danieli2017, bennet2019, smercina2018, davis2021,davis2024, mutlu2024, nashimoto2022, muller2024, muller2025, sbf_m101, carlsten2020a, carlsten2022, geha2017, mao2020, mao2024}. These surveys tend to focus on the virial volumes around known hosts, such that a clean measurement of star formation histories for isolated systems is not possible. There are compilations of known dwarfs within the Local Volume \citep[e.g.][]{karachentsev2004, karachentsev2006, karachentsev2007, karachentsev2013, karachentsev2020, jacobs2009, anand2021}, and ongoing searches for more  \citep{greco2018, zaritsky2019, tanoglidis2021, greene2022, zhang2025}, but their mass-completeness are not well-characterized to date.

We have just released a new catalog, the ELVES-Field sample, described in Carlsten et al. (submitted; hereafter Paper I). We used roughly 3,000~deg$^2$ of imaging from the Dark Energy Camera Legacy Surveys \citep{decals} with corresponding deeper imaging from the Hyper-Suprime Camera to find dwarf galaxies within 10 Mpc based on the detection of surface brightness fluctuations \citep{tonry1988}. The sample is $>50\%$ mass complete at $M_* = 10^6$~\msun\ and $>90\%$ mass complete down to $M_* = 10^{6.5}$~\msun\ out to a volume limit of 10~Mpc. In Paper I, we present the sample and investigate the luminosity and mass functions of isolated dwarfs, including a comparison with the predictions from modern cosmological simulations. In this work, we focus on the baryonic properties of the sample, including the colors, quenched fraction, and galaxy sizes.

We use the following conventions throughout this paper: all photometry is in the AB system, solar absolute magnitudes are from \citet{willmer2018}, all photometry is corrected for Galactic extinction using the $E(B-V)$ maps of \citet{sfd} and \citet{sfd2}, and effective radius, $r_e$, is taken along the major axis.

\section{Review of ELVES-Field Methodology}
\label{sec:survey}

The data sources, search algorithm, completeness tests, and distance measurements are described in detail in Paper I. Here we provide a brief overview for completeness.

\subsection{Imaging and Detection}
\label{sec:algo}
Following \citet{carlsten2022} and \citet{li2025}, we perform our search for dwarf candidates using the Southern portion of DR10 of the Legacy Surveys\footnote{\url{https://www.legacysurvey.org/dr10/description/}} \citep[LSDR10;][]{decals} conducted with the Dark Energy Camera. Because the imaging is uniform, it is much easier to quantify our completeness using this survey. However, measuring distances to dwarfs with surface brightness fluctuations \citep[SBF;][]{tonry1988, jerjen_field, jerjen_field2, cantiello2018, sbf_calib, greco2020, cantiello2024} requires higher angular resolution and deeper data, and so we only search within regions where Hyper-Suprime Camera (HSC) data are also available. %Specifically, we utilize $r$ or $i$ band HSC data from three sources: 1) the HSC Subaru Strategic Program \citep[SSP PDR3;][]{hsc, hsc_pdr3}, 2) archival HSC data reduced and made available in the HSC Legacy Archive \citep{hscla}, and 3) any remaining raw HSC data from the SMOKA archive. The HSC exposure times vary but are generally at least as deep as the `wide' layer of the HSC-SSP Survey which is 10 min in $r$ and 20 min in $i$.

The detection algorithm works in three passes with the final source list being the union of the sources found in each pass. First, we identify the brightest galaxies using the Sienna Galaxy Atlas \citep{moustakas2023}. These sources are recorded and then masked, along with bad pixels. Second, we use the \texttt{tractor} catalog to identify relatively high surface brightness ($m_g < 19$~mag or $\mu_{\rm g, eff} < 22$~mag~arcsec$^{-1}$) galaxies. These sources are recorded and then modeled and subtracted out. Very red sources with $g-r > 0.9$~mag are very likely to reside beyond the Local Volume and are removed at this stage as well. Third, once the bright galaxies have been cleaned,  we do a 2\arcsec\ smoothing and identify the faintest galaxies from this cleaned image using standard SExtractor tools.

The full list of candidates from Sienna, {\tt tractor}, and the SExtractor run are then fit with single S\'{e}rsic profiles using \texttt{pysersic}\footnote{\url{https://github.com/pysersic/pysersic}} \citep{pysersic} to the $grz$ DECaLS images. From these fits, color and size cuts are made, based on a reference sample of known Local Volume dwarfs. In brief these cuts are as follows:
\begin{itemize}
    \item $m_g < 21$ mag
    \item $g-r \in [-0.1, 0.75]$
    \item $r-z \in [-0.1, 0.5]$ 
    \item $r_e \times (1-0.5\epsilon) > 4.5\arcsec$ and $\log(r_e \times (1-0.5\epsilon)) > -0.1 m_g + 2.4$
    \item $\mu_{\rm eff, g} > 5.1 (g-r) + 19.1$~mag~arcsec$^{-1}$
    \item $n < 3$, where $n$ is the S\'{e}rsic index
\end{itemize}
We also apply a color gradient cut to remove background spiral galaxies which are often redder in the centers than in the outskirts.  To ensure these cuts do not unduly remove many true LV dwarf galaxies, we run the detection pipeline on all `bricks' ($0.25^\circ \times 0.25^\circ$ patches of LSDR10) in LSDR10 South that contain a known LV dwarf galaxy from either the Local Volume Galaxy Catalog \citep[LVGC;][]{karachentsev} or the Exploration of Local VolumE Satellites \citep[ELVES;][]{carlsten2022}, a total of 214 LV dwarfs that are brighter than $m_g < 21$ mag and with a distance measurement from either tip of the red giant branch (TRGB) or SBF.\footnote{Of these, 66 dwarfs fall into the subset of LSDR10 South that is actually surveyed in this work due to the coverage of existing HSC imaging.} Roughly 5\% of these known dwarfs are removed by the cuts, but the cuts are very effective at removing background galaxies.

After these automated cuts and a visual inspection step to remove false positives due to, e.g., Galactic cirrus, we are left with $\sim2,300$ candidate dwarfs that require distance measurements to confirm their status as LV members. We refer the reader to Paper I (Carlsten et al., submitted) for more details on the visual inspection process.

\subsection{Completeness}

We perform detailed completeness tests to quantify the completeness as a function of size, magnitude, and ultimately stellar mass. We perform two variety of tests: 1) we inject artificial dwarf galaxies into the images across a grid in size and magnitude and quantify the recovery rate and 2) we quantify how well we can recover real, known LV dwarfs in LSDR10 data. Both methods yield similar estimates to the completeness. These tests show our detection algorithm is $\gtrsim90\%$ complete down to a surface brightness of $\mu_{\rm eff, g} \sim 27$~mag~arcsec$^{-2}$ and apparent magnitude of $m_g < 21$ mag (cf. Figure 4 of Paper I)\footnote{This surface brightness sensitivity is comparable to the $\mu_{0,V}=26.5$ mag arcsec$^{-2}$ limit from ELVES \citep{carlsten2022}.}. By assuming dwarfs are uniformly distributed in the LV and obey the mass-size relation of \citet{carlsten2021a}, we can use our estimate of completeness with respect to size and magnitude to estimate completeness with respect to stellar mass. Since the mass-to-light ratio of a dwarf will depend on color, we perform this process with two different fiducial colors: $g-r=0.26$ representing a typical star-forming dwarf and $g-r=0.55$ representing a typical quenched dwarf. These are the average late-type and early-type dwarf colors, respectively, from the ELVES Survey \citep{carlsten2022}. We find that we are $>50\%$ mass complete at $M_* = 10^6$~\msun\ and $>80\%$ mass complete down to $M_* = 10^{6.5}$~\msun\ for red, quenched dwarfs. The completeness is slightly better for the blue, star-forming dwarfs due to the lower mass-to-light ratios for bluer colors. For a typical star-forming dwarf color, the detection is $>50\%$ mass complete at $M_* = 10^{5.5}$~\msun\ and $>80\%$ mass complete down to $M_* = 10^{6}$~\msun.

With an LSB-focused detection algorithm like used in ELVES-Field, there is a concern that the survey will miss very nearby dwarfs as they might be fully resolved without significant diffuse, LSB emission. However, our pipeline recovers all four known LV dwarfs within 2 Mpc in the survey footprint. Additionally, we find no significant difference in the inferred completeness if we rerun the artificial dwarf tests, injecting dwarfs at 2 Mpc so that they are mostly resolved in the LSDR10 imaging. While it is still possible that dwarfs even closer (e.g. within 1 Mpc) will be too resolved for ELVES-Field, the volume lost when considering our outer limit of 10 Mpc is insignificant. We note that complementing ELVES-Field with a resolved-star-based dwarf search using the deep HSC imaging would be a useful investigation.

\subsection{Distances and Final Local Volume Sample}
\label{sec:dist_results}
The catalog utilizes three types of distances. A small number ($\sim 1\%$) are known LV dwarfs that have tip of the red giant branch distances from \citet{karachentsev2013}. A larger fraction ($\sim 20\%$) have redshift measurements, which are converted to a distance using a flow model from \citet{kourkchi2020}, based on galaxy distances from CosmicFlows3 \citep{tully2016}\footnote{In Paper I, we quantify the accuracy of this flow model by comparing its predicted distances to existing TRGB distances. In short, we find the flow model is an unbiased estimator of distances with an uncertainty of approximately $\sim 2$ Mpc for dwarfs within 10 Mpc.}. The remaining distances are all derived from surface brightness fluctuations. 

%Surface brightness fluctuations (SBF) arise because light from a galaxy is dominated by the most massive stars \citep[e.g.,][]{tonry1988}. Thus, even when it is no longer possible to resolve individual stars in a galaxy, it is possible to measure granularity in the number of bright stars in a given resolution element. These fluctuations in brightness inversely scale with the distance. They also depend on the star formation history of a galaxy, which determines the absolute number of luminous stars. Thus, SBF distance is typically calibrated with an optical color \citep[e.g.,][]{blakeslee2009,cantiello2018}. In this work, we use either the $i$ band calibration of \citet{sbf_calib} or the $r$ band calibration of \citet{li2025}.

Based on the distance measurements, we have three categories of objects: confirmed Local Volume dwarfs, rejected background contaminants, and remaining unconfirmed/candidate Local Volume dwarfs. The rejected background contaminants have a $2 \, \sigma$ lower-bound distance beyond 10 Mpc; these are not considered further. The confirmed LV dwarfs are those that have a robust distance measurement with $2 \, \sigma$ lower-bound distance within 10 Mpc. There is some additional nuance for the LV dwarfs that are confirmed based on an SBF distance measurement. These dwarfs must have a S/N$>5$ SBF measurement as well as pass both a visual inspection and automated quality check to ensure the trustworthiness of the SBF measurement. In brief, this quality check flags dwarfs for which the SBF result sensitively depends on either the annulus size over which the measurement is taken or the range in physical scales over which the image power spectrum is fit. If the result is sensitive to either choice, it is indicative that the measured fluctuation power could be coming from spurious sources (e.g. residuals to the S\'{e}rsic fit) and not genuine SBF. We refer the reader to Section 4.3.4 of Paper I for more details. 

There is a concern when determining the quenched fraction from a dwarf sample that is created via significant use of SBF that it is biased against clumpy, star-forming dwarfs due to the difficulty of applying SBF to such systems. However, there are four important considerations indicating that this is not significantly biasing the results in this work. First, in previous SBF work \citep[e.g.][]{sbf_calib, carlsten2022, li2025} and in the SBF method demonstrations in Paper I, we have shown that SBF can accurately be applied to blue, star-forming dwarfs, particularly if the measurement focuses on the smoother outskirts of such galaxies. We refer the reader to Section 4.3.1 of Paper I for the details of this strategy in ELVES-Field. Second, the way that a clumpy light distribution in a star-forming dwarf impacts the SBF measurement is such that it generally will cause one to overestimate the fluctuation power in the image and, thus, to \textit{underestimate} the distance, making it more likely for a clumpy, star-forming dwarf to erroneously be classified a genuine LV dwarf. This is at odds with the concern that a nonzero quenched fraction at low masses is due to preferentially confirming only quenched systems with SBF. An important related point to this is that because the clumpy light generally causes an underestimate of the distance, the distance lower bounds will be conservative. While we do use SBF to actually ``confirm" LV dwarfs, it is much more heavily used in ELVES-Field to reject background contaminants by setting distance lower bounds. Third, star-forming dwarfs will more likely have redshifts, obviating the need to use SBF measurements. Finally, we have explicitly checked that the quenched fractions we find below do not actually change if we only consider a TRGB and redshift confirmed sample. The statistics drop by a factor of two, however.

Each confirmed dwarf (confirmed via SBF or otherwise) is assigned a weight to account for the chance that some galaxies will actually be outside of 10 Mpc since we impose a $2 \, \sigma$ lower bound criterion. For instance, a confirmed dwarf with a distance measurement of $10\pm1$ Mpc will have a weight of 0.5. Finally, we have the sources that remain unconfirmed. In these cases we do not have a conclusive distance measurement. Primarily this is due to the SBF S/N being too low to determine whether the galaxy falls within 10 Mpc or outside. For these, we estimate a probability of being in the Local Volume as a function of $m_g$ from the fraction of confirmed galaxies relative to the total number of candidates with conclusive distance results. Due to the high contamination rate from background galaxies, this probability is always $<5\%$.  

In the end, the sample has 95 total confirmed LV dwarfs, 1639 rejected background contaminants, and a remaining 600 unconfirmed candidates. Based on the estimated probabilities that these latter are true LV dwarfs, we expect $\sim5$ genuine LV dwarfs out of the pool of remaining unconfirmed candidates. Of the confirmed LV dwarfs, 49 are systems with previous distance measurements placing them in the LV from the LVGC \citep{karachentsev}, and the remaining 46 are either new detections here or are confirmed to be in the LV with a distance measurement for the first time by ELVES-Field. Compared to the Systematically Measuring Ultra-Diffuse Galaxies (SMUDGes) Survey \citep{zaritsky2023} which completely surveyed the LSDR10 footprint for large, LSB galaxies, 78/95 of the confirmed LV dwarfs are new detections not found in SMUDGes while 490/600 of the remaining unconfirmed candidates are newly detected here in ELVES-Field.

%\textcolor{red}{Brief explanation of flags and weights from Scott summary}

%\subsection{Photometric Measurements}

%Single-component S\'{e}rsic fitting to the DECaLS $grz$ photometry is performed for all galaxies using {\tt pysersic} \citep{pasha2023}. The errors in the S\'{e}rsic parameters are estimated from artificial galaxy injection simulations where we quantify how well we can recover galaxy properties as a function of the $g$ magnitude of the dwarf.

 %We also . 

\subsection{Isolation Criteria}
\label{sec:isolation}

As part of ELVES-Field, we present a nominal `isolated' sample. Groups are identified from \citet{kourkchi2017} based on the CosmicFlows3 distance database \citep{tully2019}. Any group with total stellar mass $M_* > 10^9$~\msun\ and within 16 Mpc is considered as a possible host for each ELVES-Field candidate. Because the distance measurements for the groups have uncertainties, it is important to consider groups beyond the nominal 10 Mpc volume\footnote{The median distance error reported by \citet{kourkchi2017} for groups with total stellar mass $M_* > 10^9$~\msun\ and within 16 Mpc is $14\%$ and, thus, a distance upper bound of 16 Mpc should conservatively include all potential hosts.}. We flag dwarfs as `isolated' if they fall outside of a projected $2 \times R_{\rm vir}$ from any of the identified groups. We emphasize that we use projected separation on the sky, and not 3D separation, when assessing a dwarf's environment to be as robust as possible to error in the dwarf's distance measurement. While TRGB distances should be precise enough ($\sim5\%$) to allow for determination of 3D separation with potential hosts, SBF and redshift-based distances are much less precise ($\sim15\%$ and $\sim20\%$, respectively) and preclude determining a 3D separation to a host. The distance measurements of the dwarfs are thus only used to confirm the dwarfs as genuine LV members or not. In this way, our determination that a given dwarf is isolated will be conservative. The lower limit on group stellar mass of $M_\star > 10^9$~\msun\ is chosen to be roughly that of the Large Magellanic Cloud as it has been found that hosts of this mass can environmentally quench their satellites \citep[e.g.][]{garling2020, jahn2022, li2025}. We caveat that, with our criterion, dwarfs considered `isolated' might indeed still be satellites of lower mass (i.e Small Magellanic Cloud-mass; $M_\star \sim 10^8$ \msun) hosts. We find similar results in the quenched fraction of isolated dwarfs, albeit with worse statistics due to fewer dwarfs being classified as `isolated', if we redo the analysis using $M_* > 10^8$~\msun\ as the lower limit on \citet{kourkchi2017} groups. 

We additionally consider an isolation criterion based on a fixed projected distance of each dwarf from any massive companion. Using the galaxy catalog of \citet{kourkchi2017}, we identify massive galaxies with $M_\star > 2.5\times10^{10}$ \msun, and flag the dwarf as a companion if it falls within 1.5 projected Mpc of the massive host, following \citet{geha2012}. We will examine the star formation properties of dwarf galaxies as a function of their distance from possible hosts below.

\section{Results}

In this paper, we focus on the baryonic properties of the ELVES-Field dwarfs. We first look at their structure in the mass-size plane and then the quenched fractions of the dwarfs as a function of their proximity to more massive galaxies.

\subsection{Mass-size Relation}
\label{sec:mass_size}
\begin{figure}
\includegraphics[width=0.48\textwidth]{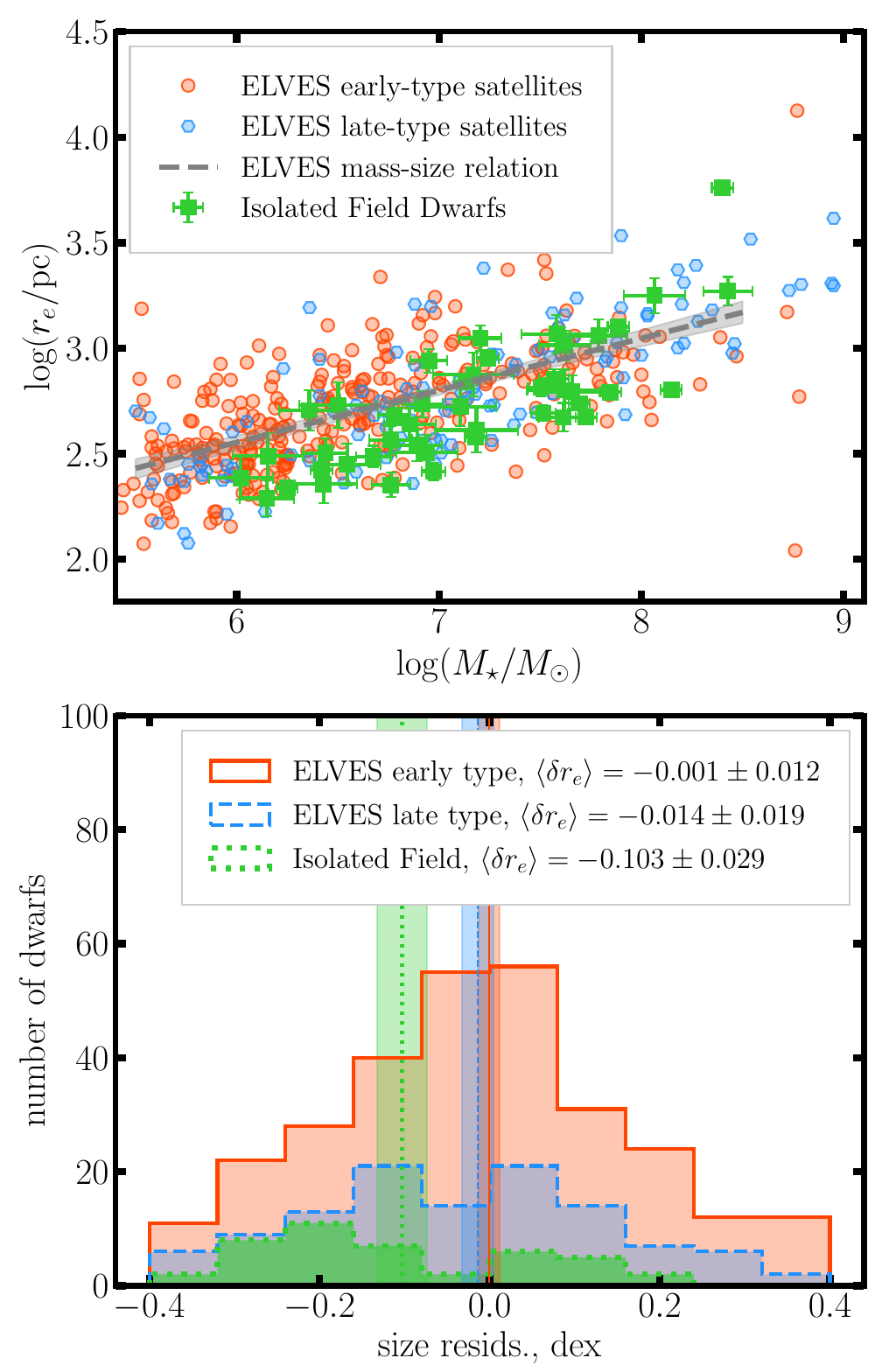}
\caption{The mass-size relation of isolated field dwarfs from ELVES-Field compared with satellite dwarfs from the ELVES Survey \citep{carlsten2022}. The top panel also shows the mass-size relation fit from ELVES in \citet{carlsten2021a} as the dashed black line. As found in ELVES, early- and late-type dwarf satellites exhibit essentially the same mass-size relation, but isolated dwarfs appear to be significantly smaller at fixed stellar mass than satellite dwarfs. The bottom panel shows histograms of the residuals from the fit mass-size relation. The vertical lines show the mean residuals for the three populations. The mean residual for the isolated dwarfs indicates they are $\sim20\%$ smaller than satellites dwarfs at the same stellar mass. }
\label{fig:msr}
\end{figure}

Figure \ref{fig:msr} shows the mass-size relation for isolated dwarfs detected here compared to satellite dwarfs from the ELVES Survey. We only show distance-confirmed dwarfs and consider only dwarfs that are outside of $2\times R_{\mathrm{vir}}$ projected from any $M_\star >10^9$ \msun~ groups as `isolated'. From the luminosities and colors, we calculate stellar masses based on the color-mass relations for dwarf galaxies presented in \citet{delosreyes2024}. As found in \citet{carlsten2021a}, late- and early-type satellite dwarfs exhibit essentially the same mass-size relation, but isolated dwarfs appear to be significantly smaller at fixed stellar mass. The bottom panel of Figure \ref{fig:msr} shows the histograms of residuals from the mass-size relation fit of \citet{carlsten2021a}, indicating that isolated dwarfs from ELVES-Field are $\sim20\%$ smaller at fixed mass than the satellite dwarfs of the ELVES Survey. Very recently, \citet{asali2025} found a similar size offset at higher dwarf masses comparing satellites from the Satellites Around Galactic Analogs Survey \citep[SAGA;][]{mao2024} with field dwarfs from SDSS, among other sources.

This result was hinted at in \citet{carlsten2021a} where we found that isolated dwarfs from the Local Volume Galaxy Catalog \citep[LVGC;][]{karachentsev} were also smaller at fixed mass than ELVES satellite dwarfs. However, a strong conclusion could not be drawn in that work, since the completeness of the LVGC is not established and could feasibly be missing large, LSB dwarfs. In the current work, however, ELVES and ELVES-Field have essentially the same surface brightness sensitivity of $\mu_{\rm eff, g} \sim 27$~mag~arcsec$^{-2}$ as shown through injection tests of artificial dwarfs. ELVES-Field has an explicit size cut on circularized effective radius of 4.5\arcs. While ELVES did not have an explicit size cut, the required number of pixels required for detection placed roughly a $r_e > 3$\arcs~ requirement, smaller than the ELVES-Field criterion.

We note that the S\'{e}rsic fitting methodology used here is somewhat different compared to that used in the original ELVES Survey, which could have a significant impact on this comparison. However, when comparing the sizes reported by the pipeline used here for ELVES and LVGC dwarfs that fall in the LSDR10 footprint, we find good agreement with the sizes from \citet{carlsten2021a} without discernible bias. In particular, as part of the completeness tests, we run the ELVES-Field pipeline over all known LV dwarfs in the LSDR10-South footprint from ELVES and the LVGC. There are 203 such dwarfs with existing photometry from \citet{carlsten2021a} that pass the ELVES-Field photometric cuts (cf. \S\ref{sec:algo}). The median residual in $\log(r_e)$ between the ELVES-Field photometry and that from \citet{carlsten2021a} is $-0.004$ dex with mean of $0.013$ dex. The median reported uncertainty in $\log(r_e)$ from \citet{carlsten2021a} is $0.03$ dex.

This observation could indicate that satellite dwarfs all experience some tidal processing effect that increases their sizes \citep[e.g.][]{mayer2001a, penarrubia2008}. However, this is surprising, because late-type satellites are thought to have only recently fallen into the halo of their host galaxies as the quenching mechanisms at these low masses ($M_\star\sim10^6-10^8$ \msun) should operate quickly \citep[e.g.][]{fillingham2015, akins2021, greene2023}. An interesting avenue for future work would be to check if this trend exists for MW satellites compared to nearby Local Group isolated dwarfs (e.g. Tucana), but we caution that consistent size measurement methodology will likely be required to see a difference of this size. We leave further exploration as to why isolated dwarfs appear to be smaller than satellite dwarfs at fixed mass (even late-type satellites) to future work. 

% \textcolor{red}{Any chance distance uncertainty could play a role here?} -- no because even just the trgb subsample shows a meaningful offset to smaller sizes 

\subsection{Star-Forming Properties}
\label{sec:quenching}
We now present our findings on the star-forming properties of dwarfs in the ELVES-Field sample. We first show the colors and NUV properties of detected dwarfs and provide a criterion for deciding if a dwarf is `quenched'. Then we present results on the quenched fraction of isolated field dwarfs.

\subsubsection{Dwarf Colors}
\label{sec:colors}
\begin{figure}
\includegraphics[width=0.48\textwidth]{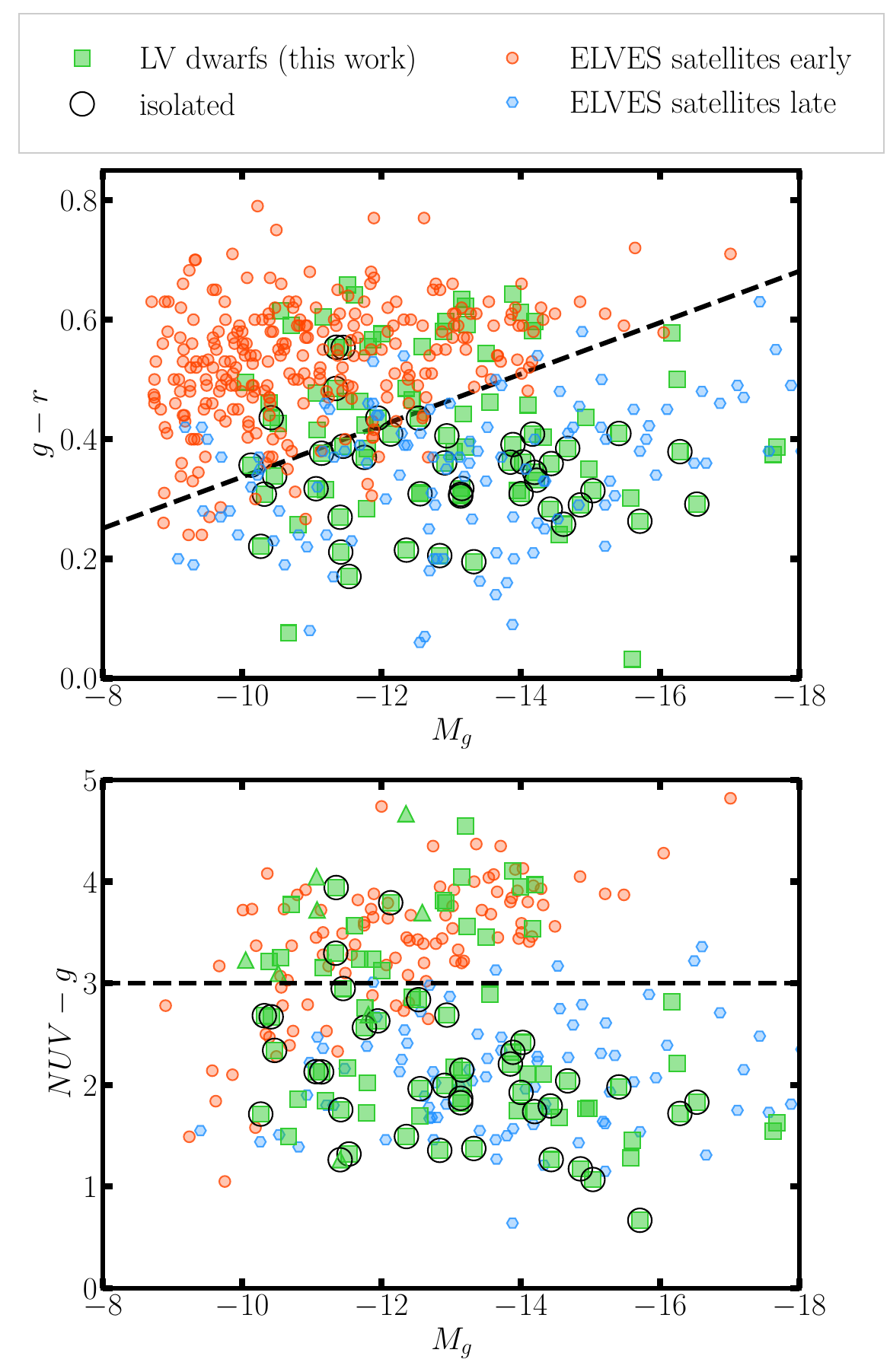}
\caption{The optical and NUV colors of confirmed LV dwarfs in ELVES-Field. Points enclosed in black circles represent dwarfs that are `isolated' according to our fiducial definition from \S\ref{sec:isolation}. Also shown for reference are satellite dwarfs from the ELVES Survey \citep{carlsten2022}, split by morphology. Dwarfs not detected in GALEX NUV data are shown as upward arrows at their $2\sigma$ lower limits of $NUV-g$ color. The dashed black lines show the thresholds in color that we use to label dwarfs as `quenched' or `star-forming'. When available, we use the threshold in $NUV-g$ with higher priority than that in optical $g-r$.}
\label{fig:color_mag}
\end{figure}

Figure \ref{fig:color_mag} shows both the optical and NUV colors of the confirmed LV dwarfs in ELVES-Field. The NUV photometry comes from aperture-based photometry using archival GALEX imaging, where available. The top panel shows the $g-r$ color vs. the $g$ band luminosity while the bottom panel shows the $NUV-g$ colors. For LV dwarfs without a S/N$>2$ detection, we plot the $2\sigma$ lower bound to their $NUV-g$ color as an upward triangles. Points enclosed in a circle represent dwarfs that are isolated according to our fiducial definition in \S\ref{sec:isolation}. For reference, we also plot the ELVES satellites from \citet{carlsten2022} split on their visual morphology classification (early-type or late-type). Only ELVES satellites with S/N$>2$ GALEX NUV detections are plotted in the bottom panel.

Like with the original ELVES Survey, because ELVES-Field is a photometric survey, determining the quenched fraction of dwarfs is not straightforward. Without spectroscopic follow-up data, we rely on dwarf color as the best indicator of star-forming activity. NUV information from GALEX when available is particularly powerful, as UV emission is especially sensitive to recent star formation in dwarfs \citep{lee2009, karachentsev2013_sfr}. In \citet[][see Figure 10]{carlsten2022}, we found that ELVES satellites and  satellites from the SAGA Survey \citep{geha2017, mao2020, mao2024} were divided at roughly $NUV-g=3$ between visually classified early- and late-type dwarfs (in the case of ELVES) and between quenched and star-forming (in the case of SAGA which used H$\alpha$ as an indicator of star-formation). Here, we take this threshold to distinguish between quenched and star-forming field LV dwarfs. We do not attempt to convert the NUV fluxes into physical star-formation rates \citep[e.g.][]{iglesias2006} since those calibrations are designed for higher metallicity, more massive galaxies and are liable to yield spurious estimates for dwarfs \citep[see discussion in][]{greene2023}.

For dwarfs without GALEX coverage or weak upper bounds on the NUV flux, we rely on optical color to determine their quenched status. In \citet{carlsten2021a}, we found that the line $g-i=-0.067\times M_V - 0.23$ fairly cleanly separates early- from late-type dwarfs. \citet{font2021b} confirmed that this line did a fair job of distinguishing star-forming from quenched satellites in the ARTEMIS simulations. \citet{mao2024} converted this line into $M_r - (g-r)$ space and showed that it did a good job of separating star-forming from quenched satellites (as defined by the presence or not of H$\alpha$ emission) with around $\sim90\%$ purity and completeness. We further convert the line from \citet{mao2024} into $M_g - (g-r)$ space, finding $g-r = -0.0430 M_g - 0.0929$, which is shown in Figure \ref{fig:color_mag}. Based on the success that this color-magnitude threshold has in separating quenched from star-forming dwarfs in simulations and spectroscopic observations, we use it to label dwarfs as quenched or not in the absence of GALEX NUV constraints. For unconfirmed, candidate LV dwarfs, we rely solely on this optical color-magnitude threshold as the large number of candidates makes it prohibitive to search for GALEX data. These unconfirmed candidates are not plotted in Figure 2 but are incorporated in the calculation of the quenched fraction below.

\subsubsection{Quenched Fractions}
\label{sec:qf}
Using the color-based criteria for determining whether dwarfs are quenched or not (Figure \ref{fig:color_mag}), in this section, we present the quenched fractions of ELVES-Field dwarfs as both a function of dwarf stellar mass and distance to the nearest massive companion.

\begin{figure}
\includegraphics[width=0.48\textwidth]{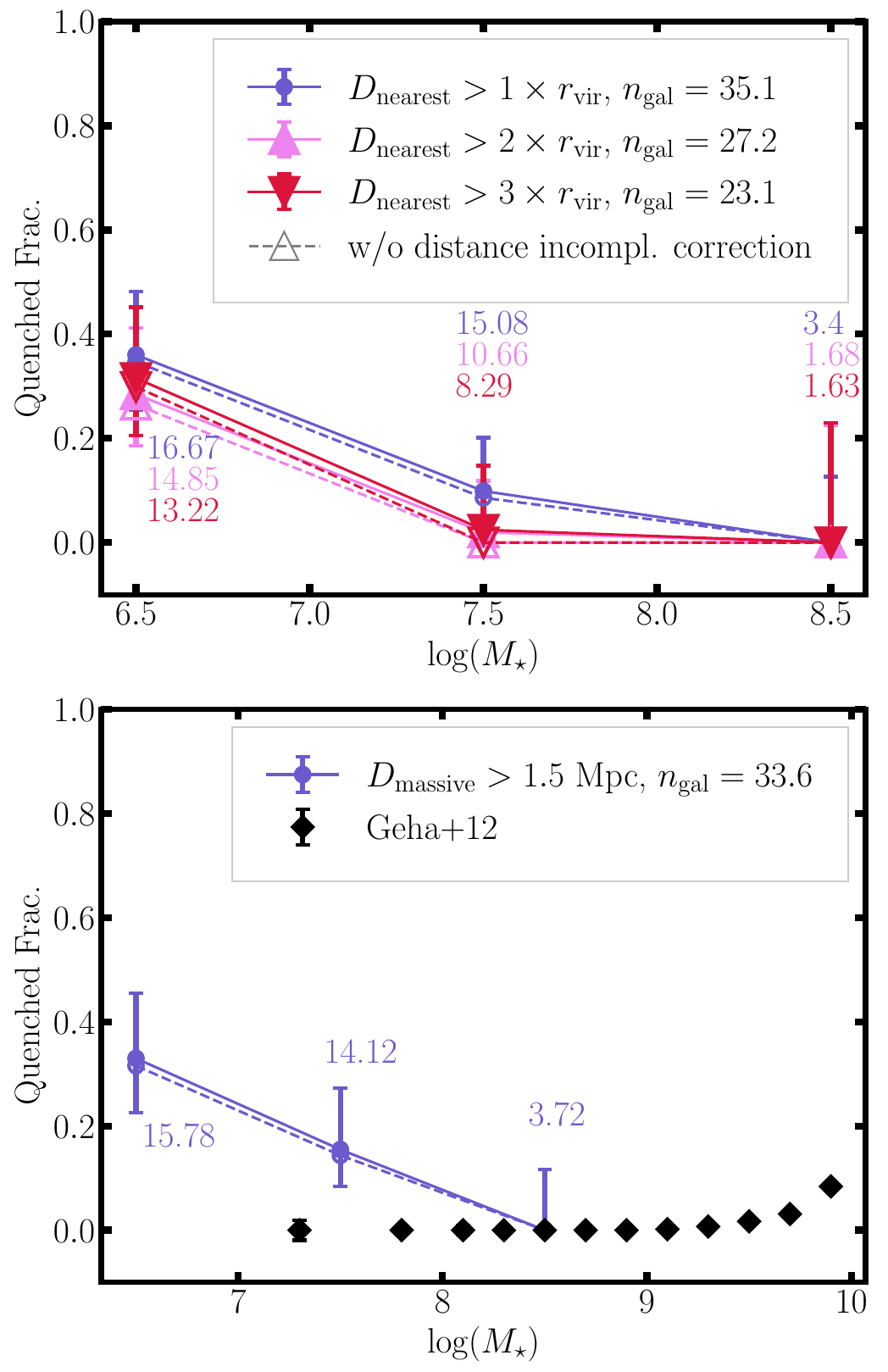}
\caption{The quenched fraction of isolated dwarfs versus dwarf stellar mass. Dwarfs are inferred to be quenched or star-forming based on their $NUV-g$ or $g-r$ color. The curves represent different criteria for a dwarf to be considered `isolated'. The top panel shows criteria based on dwarfs being outside of $1\times,~ 2\times,$ or $3\times$ the projected virial radii of any massive, $M_\star > 10^9$ \msun~ group. The bottom panel shows the criterion used in \citet{geha2012} where dwarfs are required to be $>1.5$ Mpc projected from any $M_\star > 2.5\times 10^{10}$ \msun~ galaxy. The numbers listed in the legends and near each data point show the average number of galaxies contributing to each mass bin after we account for unconfirmed dwarfs without distance measurements. Dashed curves portray the results if this incompleteness correction is not performed, showing that it has a minor effect. Errorbars show the $1\sigma$ Bayesian confidence region. Due to the small number of isolated dwarfs found in the ELVES-Field footprint, the statistics are rather poor. However, we find that high mass dwarfs are essentially all star-forming and find evidence that quenched fractions increase to $\sim30\%$ for $M_\star < 10^7$ \msun, irregardless of isolation criterion used. The bottom panel shows the results from \citet{geha2012} for reference at higher masses.}
\label{fig:qf_mstar}
\end{figure}

Figure \ref{fig:qf_mstar} shows the quenched fraction of isolated dwarfs as a function of dwarf stellar mass. The top panel shows the results for different isolation criteria requiring the dwarfs to be outside of a projected $1\times R_{\mathrm{vir}}$, $2\times R_{\mathrm{vir}}$, or $3\times R_{\mathrm{vir}}$ from any massive, $M_\star > 10^9$ \msun, \citet{kourkchi2017} group, as described in \S\ref{sec:isolation}. The bottom panel shows the results for an isolation criterion similar to that used in \citet{geha2012} where dwarfs are required to be more than a projected 1.5 Mpc from any $M_\star > 2.5\times 10^{10}$ \msun~ galaxy from \citet{kourkchi2017}. In each case, the number of dwarfs satisfying the isolation criterion is shown in the legend. Each confirmed dwarf contributes to the average according to its `confirmed weight' (\S\ref{sec:dist_results}), and we include the contribution of unconfirmed dwarfs without distance measurements by stochastically including them according to their estimated probabilities of being real LV dwarfs. For these unconfirmed dwarfs, their quenched/star-forming status is determined from $g-r$ color as we do not analyze any GALEX data for them. This incompleteness correction is the reason why the numbers of included dwarfs are non-integral. The dashed curves in Figure \ref{fig:qf_mstar} show the quenched fractions if this incompleteness correction is not made, demonstrating that it has a fairly minor effect.  The error bars show the Bayesian ($1\sigma$) confidence region for the quenched fraction, treating it as a binomial proportion parameter with an uninformative Jeffreys prior\footnote{\url{https://docs.astropy.org/en/stable/api/astropy.stats.binom_conf_interval.html}.}.

Unfortunately with the limited area footprint of ELVES-Field, there are only a few dozen isolated dwarfs, and the statistics on the quenched fraction are rather poor. With that said, we do find that the quenched fraction is low at higher ($M_\star \gtrsim 10^7$ \msun) dwarf masses, and there is evidence that it increases at lower masses. The $10^6 < M_\star < 10^7$ \msun~ bin for the fiducial $D_\mathrm{nearest} > 2\times R_{\mathrm{vir}}$ isolation criterion has $\sim 4 / 15$ dwarfs identified as `quenched'. The fact that this is non-zero is still an interesting conclusion, however, and is in-line with recent one-off discoveries of isolated quenched dwarfs \citep[e.g.][]{makarov2012, karachentsev2015_kks3, polzin2021, li2024_hedgehog, sand2024}. Interestingly, the quenched fractions are quite similar across all of the isolation criteria. We emphasize that the isolation criteria are all based on projected (i.e. on-the-sky) separation and not 3D separation and are, thus, robust to the measurement errors in the distance to the LV dwarfs.

A robust comparison to theoretical predictions for the isolated dwarf quenched fraction is difficult as different definitions of `isolated' abound in the literature. However, as mentioned in the introduction, several recent theoretical works (both simulation-based and semi-analytic) predict appreciable ($\sim30\%$) quenched fractions for dwarfs at the low masses probed here ($M_\star \sim 10^6-10^7$ \msun). \citet{wang2024} developed a semi-analytic model of dwarf galaxy evolution and found that, in order to explain the high quenched fraction of SAGA satellites, isolated field dwarfs must reach $\sim30\%$ quenched fractions by $M_\star\sim10^{6.5}$ \msun. Here, `isolated' means the same criterion used in \citet{geha2012} which is what we use in the bottom panel of Figure \ref{fig:qf_mstar}. Using full hydrodynamic simulations, \citet{christensen2024} predict a similar quenched fraction for field dwarfs where `field' in this case means `non-satellite'.

The mechanism(s) behind the quenching of the isolated field dwarfs we find in ELVES-Field remain unclear. The stellar mass of $M_\star \sim 10^{6.5}$ \msun~ of the quenched dwarfs is higher than what is generally considered the threshold for quenching due to the UV background during reionization \citep[e.g.][]{bullock2000, bovill2011}, although it is still feasible that reionization plays a role. Alternatively, the field dwarfs could be quenched by ram-pressure removal of their gas as they pass through filaments and sheets in the cosmic web \citep{benitez2013, pasha2023, benavides2025, luber2025}. However, \citet{bidaran2025} reports the discovery of some quenched, isolated dwarf galaxies in cosmic voids, far from cosmic web filaments or sheets.

Additionally, the quenching could be self-quenching from stellar feedback within the dwarf itself \citep{samuel2022}. In the Feedback In Realistic Environments-2 hydrodynamic simulations, \citet{samuel2022} find a substantial ($>40\%$) quenched fraction for `central' dwarf galaxies with $M_\star \sim 10^6-10^7$ \msun~ that are 1-2 Mpc away from a MW-sized host and have never been within the virial radius of a more massive halo and attribute this to a combination of reionization and stellar feedback effects. Unfortunately they caveat that the resolution of the simulations might also play a role and cause some dwarfs to artificially quench. \citet{samuel2022} also highlight the effect of processing in low mass groups. While our sample of isolated field dwarfs are, by definition, not in groups with galaxies with $M_\star > 10^9$ \msun, they could be satellites of $M_\star \sim 10^8$ \msun~ (i.e. Small Magellanic Cloud-mass) hosts \citep{li2025} which might be able to environmentally quench their satellites. This would be a surprising result, however, as hosts of these masses are not expected to have hot gas halos \citep{correa2018} and thus would not be able to ram-pressure strip their satellites like it is thought that MW-mass hosts do. Indeed, a handful of the confirmed dwarfs flagged as `isolated' here appear to likely be satellites of a nearby SMC-massed host, including one of the quenched isolated dwarfs. These dwarfs overlap with the ELVES-Dwarf sample \citep{li2025} and will be discussed further in the context of that project. 

Finally, the quenched isolated dwarfs could simply be `backsplash' galaxies that previously had passed through the virial volume of a massive host, even if they now reside far outside \citep{bhattacharyya2025}. \citet{teyssier2012} estimated these constitute $\gtrsim15\%$ of halos even out to 1.5 Mpc from a MW-like host. Similarly, \citet{buck2019} estimated $\gtrsim 40\%$ of subhalos are backsplash out to $2\times R_\mathrm{vir}$. Velocity measurements of the quenched dwarfs in our sample would help test this by facilitating simple kinematic modeling relative to nearby putative hosts. In the Local Group, \citet{bennet2025} did detailed kinematic modeling to show that the relatively isolated quenched dwarf Tucana is very likely not a backsplash galaxy from either the MW or M31. 

\begin{figure}
\includegraphics[width=0.48\textwidth]{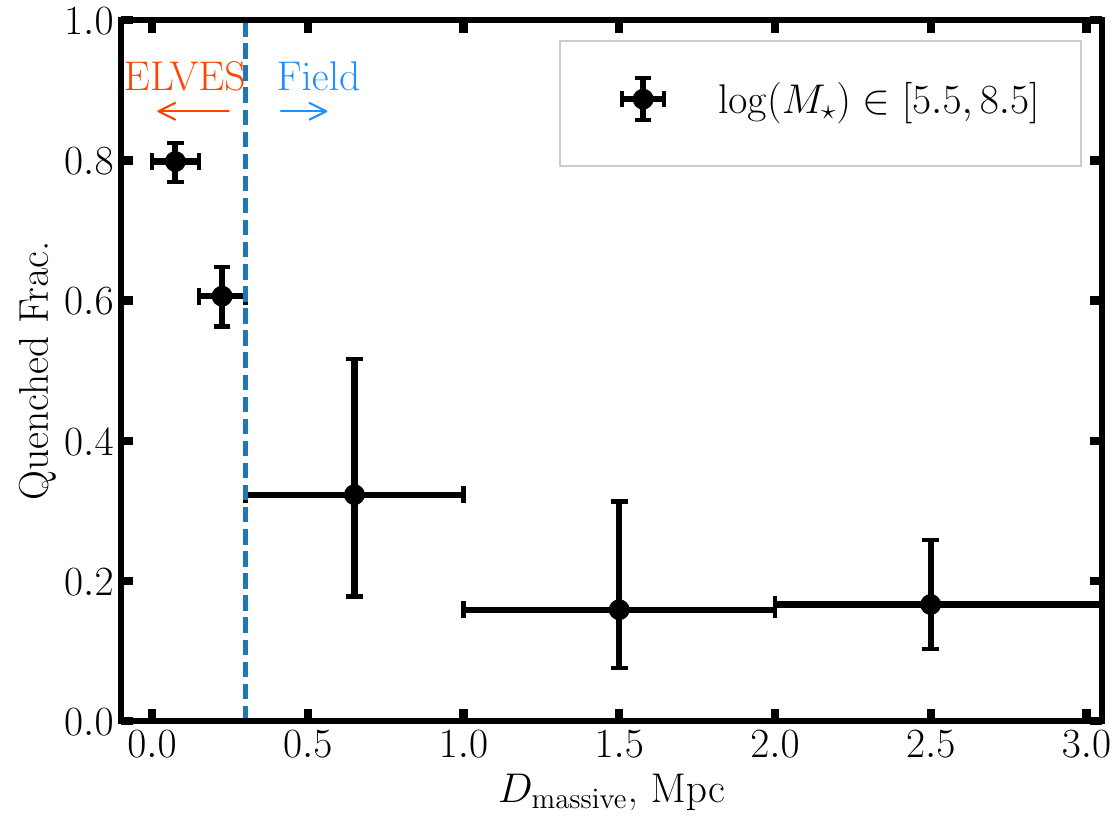}
\caption{The average quenched fraction of dwarfs in the mass range $10^{5.5} < M_\star < 10^{8.5}$ \msun~ as a function of the projected separation of the dwarf to the nearest massive ($M_\star > 1\times 10^{10}$ \msun) host galaxy. In assigning this separation, we take the minimum projected distance of each dwarf to the massive galaxies in \citet{kourkchi2017} that are at a distance consistent within $2\sigma$ to that of the dwarf. We find a non-zero quenched fraction even out to $2-3$ Mpc. To complement the field dwarf sample, within 300 kpc we plot the results of the ELVES Survey.}
\label{fig:fq_dist}
\end{figure}

To investigate the backsplash possibility, we show in Figure \ref{fig:fq_dist} the average quenched fraction of dwarfs as a function of their projected separation to the nearest massive, $M_\star > 1\times 10^{10}$ \msun, potential host galaxy. This mass threshold is chosen to roughly correspond to the lower mass limit for ELVES `MW-analog' host galaxies. For this plot, we only include fully confirmed LV dwarfs. In assigning the distance to the nearest massive host galaxy, we simply take the minimum projected separation for a given dwarf to each massive galaxy from \citet{kourkchi2017} within a distance within $2\sigma$ of the distance of that dwarf. We also include the quenched fraction for satellites in the ELVES Survey which probe separations out to 300 kpc from their massive hosts. As found in \citet{greene2023}, there is a trend whereby ELVES satellites closer in projection to their hosts have somewhat higher quenched fractions. The ELVES-Field dwarfs probe the quenched fraction out to several Mpc in separation and still have a non-zero quenched fraction at $2-3$ Mpc of projected separation. This fact combined with the result in Figure \ref{fig:qf_mstar} where field dwarfs beyond $3\times R_\mathrm{vir}$ of any massive galaxy can also be quenched would indicate that not all of the quenched dwarfs are backsplash and that some other mechanism is at work. A more direct comparison with simulations and likely a larger observational sample size will be required to fully elucidate the origin of these quenched field dwarfs. Additionally, space-based resolved star observations would allow for detailed reconstruction of their star formation histories, providing clues to what physical effects could be responsible \citep[e.g.][]{mutlu2025}.

\section{Discussion and Summary}
\label{sec:conclusions}
ELVES-Field is a survey of the Local Volume dwarf population over $\sim3,000$ square degrees. In Paper I, we provide details of the survey design, detection algorithm, and luminosity and stellar mass functions of the detected dwarfs. Paper I also establishes the completeness of the sample down to $\mu_{\rm eff, g} \approx 27$~mag~arcsec$^{-2}$ in surface brightness and $M_\star \approx 10^6$ \msun~ in stellar mass. In the current paper, we focus on the baryonic properties of the dwarfs, including their mass-size relation and star-forming properties.

In particular, we focus on isolated field dwarfs. We consider different criteria for selecting isolated dwarfs but primarily select isolated dwarfs as those that are outside of a projected  $2\times R_{\mathrm{vir}}$ from any massive group with $M_\star > 10^9$ \msun~ within 16 Mpc from the group catalog of \citet{kourkchi2017}. 

Figure \ref{fig:msr} shows that isolated dwarfs from ELVES-Field are significantly smaller at fixed stellar mass than satellite dwarfs from the ELVES Survey. The difference is $\sim 20\%$. This result has been hinted at in previous research \citep[e.g.][]{carlsten2021a, li2024_hedgehog}, but required the completeness of the ELVES-Field sample to show conclusively. This is a somewhat surprising result given the current understanding that late-type satellites will have only recently fallen into their host halos and thus should be quite similar to field late-type dwarfs. 

We also investigate the quenched fraction of field dwarfs. We distinguish between star-forming and quenched dwarfs using optical and NUV colors (cf. Figure \ref{fig:color_mag}). Figure \ref{fig:qf_mstar} shows the estimated quenched fraction as a function of dwarf stellar mass using different isolation criteria. We find that at higher dwarf masses ($M_\star \gtrsim 10^7$ \msun), nearly all field galaxies are star-forming, in agreement with previous results \citep[e.g.][]{geha2012, kadofong2025}. At lower masses, we find evidence of a significant ($\sim 30\%$) quenched population. The mechanism(s) responsible for the quenching of these isolated dwarfs is unclear. It is possible that reionization plays a role \citep{kim2024}, the dwarfs are quenched via ram pressure stripping as they pass through filaments in the cosmic web \citep{benitez2013}, the dwarfs self-quench through internal stellar feedback \citep{samuel2022}, or the dwarfs are simply `backsplash` galaxies that have previously traveled nearby a massive host some time in the past \citep{bhattacharyya2025}. Based on our definition of `isolated', it is also possible that some of the `isolated' dwarfs are, in fact, satellites of lower mass (e.g. $M_\star \sim 10^8$ \msun) hosts, and these hosts are capable of environmental quenching \citep[e.g.][]{garling2024}.

The statistics of ELVES-Field is, unfortunately, rather poor and the quenched population corresponds to $\sim 4 / 15$ galaxies in the sample with mass $10^6 < M_\star < 10^7$ \msun~. To fully investigate what physical process is responsible for the quenching, surveys over a larger area of the sky will be needed to obtain better statistics. ELVES-Field is based on deep, ground-based data from HSC/Subaru, and thus a very similar methodology (including the heavy use of SBF) can be applied to the much wider field Vera Rubin Observatory's Legacy Survey of Space and Time in the coming years. This will increase the sample size $5-10\times$. Additionally, Euclid \citep[e.g.][]{marleau2025, hunt2025} and the Roman Space Telescope will facilitate LV dwarf searches using only tip of the red giant branch measurements. With these space-based surveys, it will also be possible to infer star formation histories from resolved star measurements which will greatly help elucidate the quenching processes at work in isolated, low-mass galaxies.

\section*{Acknowledgements}
This work was supported by NSF AST-2006340. J.L. and J.E.G. gratefully acknowledge support from the NSF grant AST-2506292. J.L. acknowledges support from the Charlotte Elizabeth Procter Fellowship at Princeton University. 

The Legacy Surveys consist of three individual and complementary projects: the Dark Energy Camera Legacy Survey (DECaLS; Proposal ID \#2014B-0404; PIs: David Schlegel and Arjun Dey), the Beijing-Arizona Sky Survey (BASS; NOAO Prop. ID \#2015A-0801; PIs: Zhou Xu and Xiaohui Fan), and the Mayall z-band Legacy Survey (MzLS; Prop. ID \#2016A-0453; PI: Arjun Dey). DECaLS, BASS and MzLS together include data obtained, respectively, at the Blanco telescope, Cerro Tololo Inter-American Observatory, NSF’s NOIRLab; the Bok telescope, Steward Observatory, University of Arizona; and the Mayall telescope, Kitt Peak National Observatory, NOIRLab. The Legacy Surveys project is honored to be permitted to conduct astronomical research on Iolkam Du’ag (Kitt Peak), a mountain with particular significance to the Tohono O’odham Nation.

NOIRLab is operated by the Association of Universities for Research in Astronomy (AURA) under a cooperative agreement with the National Science Foundation.

This project used data obtained with the Dark Energy Camera (DECam), which was constructed by the Dark Energy Survey (DES) collaboration. Funding for the DES Projects has been provided by the U.S. Department of Energy, the U.S. National Science Foundation, the Ministry of Science and Education of Spain, the Science and Technology Facilities Council of the United Kingdom, the Higher Education Funding Council for England, the National Center for Supercomputing Applications at the University of Illinois at Urbana-Champaign, the Kavli Institute of Cosmological Physics at the University of Chicago, Center for Cosmology and Astro-Particle Physics at the Ohio State University, the Mitchell Institute for Fundamental Physics and Astronomy at Texas A\&M University, Financiadora de Estudos e Projetos, Fundacao Carlos Chagas Filho de Amparo, Financiadora de Estudos e Projetos, Fundacao Carlos Chagas Filho de Amparo a Pesquisa do Estado do Rio de Janeiro, Conselho Nacional de Desenvolvimento Cientifico e Tecnologico and the Ministerio da Ciencia, Tecnologia e Inovacao, the Deutsche Forschungsgemeinschaft and the Collaborating Institutions in the Dark Energy Survey. The Collaborating Institutions are Argonne National Laboratory, the University of California at Santa Cruz, the University of Cambridge, Centro de Investigaciones Energeticas, Medioambientales y Tecnologicas-Madrid, the University of Chicago, University College London, the DES-Brazil Consortium, the University of Edinburgh, the Eidgenossische Technische Hochschule (ETH) Zurich, Fermi National Accelerator Laboratory, the University of Illinois at Urbana-Champaign, the Institut de Ciencies de l’Espai (IEEC/CSIC), the Institut de Fisica d’Altes Energies, Lawrence Berkeley National Laboratory, the Ludwig Maximilians Universitat Munchen and the associated Excellence Cluster Universe, the University of Michigan, NSF’s NOIRLab, the University of Nottingham, the Ohio State University, the University of Pennsylvania, the University of Portsmouth, SLAC National Accelerator Laboratory, Stanford University, the University of Sussex, and Texas A\&M University.

BASS is a key project of the Telescope Access Program (TAP), which has been funded by the National Astronomical Observatories of China, the Chinese Academy of Sciences (the Strategic Priority Research Program “The Emergence of Cosmological Structures” Grant \# XDB09000000), and the Special Fund for Astronomy from the Ministry of Finance. The BASS is also supported by the External Cooperation Program of Chinese Academy of Sciences (Grant \# 114A11KYSB20160057), and Chinese National Natural Science Foundation (Grant \# 11433005).

The Legacy Survey team makes use of data products from the Near-Earth Object Wide-field Infrared Survey Explorer (NEOWISE), which is a project of the Jet Propulsion Laboratory/California Institute of Technology. NEOWISE is funded by the National Aeronautics and Space Administration.

The Legacy Surveys imaging of the DESI footprint is supported by the Director, Office of Science, Office of High Energy Physics of the U.S. Department of Energy under Contract No. DE-AC02-05CH1123, by the National Energy Research Scientific Computing Center, a DOE Office of Science User Facility under the same contract; and by the U.S. National Science Foundation, Division of Astronomical Sciences under Contract No. AST-0950945 to NOAO.

The Hyper Suprime-Cam (HSC) collaboration includes the astronomical communities of Japan and Taiwan, and Princeton University. The HSC instrumentation and software were developed by the National Astronomical Observatory of Japan (NAOJ), the Kavli Institute for the Physics and Mathematics of the Universe (Kavli IPMU), the University of Tokyo, the High Energy Accelerator Research Organization (KEK), the Academia Sinica Institute for Astronomy and Astrophysics in Taiwan (ASIAA), and Princeton University. Funding was contributed by the FIRST program from the Japanese Cabinet Office, the Ministry of Education, Culture, Sports, Science and Technology (MEXT), the Japan Society for the Promotion of Science (JSPS), Japan Science and Technology Agency (JST), the Toray Science Foundation, NAOJ, Kavli IPMU, KEK, ASIAA, and Princeton University. 

This paper makes use of software developed for Vera C. Rubin Observatory. We thank the Rubin Observatory for making their code available as free software at http://pipelines.lsst.io/.

This paper is based on data collected at the Subaru Telescope and retrieved from the HSC data archive system, which is operated by the Subaru Telescope and Astronomy Data Center (ADC) at NAOJ. Data analysis was in part carried out with the cooperation of Center for Computational Astrophysics (CfCA), NAOJ. We are honored and grateful for the opportunity of observing the Universe from Maunakea, which has the cultural, historical and natural significance in Hawaii.

\software{ \texttt{astropy} \citep{astropy, bradley2020}, \texttt{sep} \citep{sep}, \texttt{imfit} \citep{imfit}, \texttt{SWarp} \citep{swarp}, \texttt{Scamp} \citep{scamp}, \texttt{SExtractor} \citep{sextractor}, \texttt{astrometry.net} \citep{astrometry_net}   } 

\bibliographystyle{aasjournal}
\bibliography{calib}

\end{CJK*}
\end{document}